# Digital Twinning Remote Laboratories for Online Practical Learning

C. Palmer[a,b,*], B. Roullier[b, *], M. Aamir [a,b], F. McQuade[b], Leonardo Stella[a], A. Anjum[c] and U. Diala[a]

[a] *University of Derby, Kedleston Road, Derby, DE22 1GB*

[b] *Bloc Digital, 2nd Floor, Enterprise Centre, Bridge St, Derby, DE1 3LD, UK*

[c] *University of Leicester, University Road, Leicester, LE1 7RH*

# Digital Twinning Remote Laboratories for Online Practical Learning

**Abstract.** The COVID19 pandemic has demonstrated a need for remote learning and virtual learning applications such as virtual reality (VR) and tablet-based solutions. Creating complex learning scenarios by developers is highly time-consuming and can take over a year. It is also costly to employ teams of system analysts, developers and 3D artists. There is a requirement to provide a simple method to enable lecturers to create their own content for their laboratory tutorials. Research has been undertaken into developing generic models to enable the semi-automatic creation of a virtual learning tools for subjects that require practical interactions with the lab resources. In addition to the system for creating digital twins, a case study describing the creation of a virtual learning application for an electrical laboratory tutorial is presented, demonstrating the feasibility of this approach.

**Keywords.** Virtual Reality Learning, Digital Twin, Remote Learning

## 1. Introduction

As a response to the current pandemic crisis interest has been generated in remote learning and virtual learning applications such as virtual reality (VR) and tablet-based solutions. The Guardian found that it is widely recognised that online learning is here to stay (Guardian, 2020). The Harvard Business Review (Gallagher and Palmer, 2020) states that “This moment is likely to be remembered as a critical turning point between the “time before,” when analogue on-campus degree-focused learning was the default, to the “time after,” when digital, online, career-focused learning became the fulcrum of competition between institutions". To achieve good quality career-focused learning, there is a need for university science and technology courses to offer practical experience replicating industrial equipment and scale. Even in ordinary times laboratory provision is costly and may not supply the diversity, complexity or realistic indication of the dimensions encountered in industry. Due to limited provision, and health and safety requirements, student access to laboratories may be restricted, providing an educational barrier to groups such as part-time and commuting students, and those with limited mobility. In the UK there is a move towards more students commuting to campus to save money (Guardian, 2021).

Virtual Reality has a heavy background in its potential usage for education. It allows a user to learn through interaction with the virtual world, allowing learning from experience to take place (Pantelidis, 2009; Kolb; 1984). VR learning offers a safe and realistic environment with wider accessibility than physical laboratories. By enabling users to train on multiple makes and models of equipment VR learning is able to supply a wide range of experience VR learning can accommodate individual differences in terms of learning style (Chen et al., 2005) and enables students to repeat experiments in order to gain proficiency. Childs et al. (2021) have found that Augmented Reality (AR) "and VR have also been shown to improve student engagement and motivation in the classroom, as immersive lectures are much more interesting and comprehensive than lectures presented on a 2-dimensional screen”. Childs et al. also found that VR and AR technologies can overcome distance education issues such as decreased amount of social interaction, lack of discourse and absence of non-verbal cues within these environments, as they encourage communication more effectively than traditional audio or video methods. Recently, an immersive virtual reality-based model has been proposed, namely CAMIL, which links the learning in an immersive virtual reality environment to six cognitive factors comprising: interest, motivation, cognitive load, self-efficacy, self-regulation and embodiment. Self-efficacy refers to an individual's perceived capabilities for learning or performing actions. Self-regulation is the ability to maintain attention focus in the presence of distractions. In the context of VR, embodiment refers to the experience of owning a virtual body and can be influenced by the external appearance of the body, the ability to control the body's actions and the possibility to perceive events affecting the body (Makransky & Petersen, 2021). Chau et al. (2021) discuss student attitudes which impact on self-directed learning.

It should be noted that the term VR refers to the complete immersion of the user in a 3D digital environment using a head-mounted display such as the Oculus Quest or Vive Index. The authors intend to limit the use of the term VR to fully immersive within this paper. Augmented reality enhances the real-world with images and text, via devices such as smartphones, tablets and AR glasses (e.g., Hololens).

A Virtual World learning solution requires the development of a virtual environment, e.g., a laboratory, containing a VR replica of the workshop/laboratory, equipment, equipment layout and attribute data, the laboratory procedures to be followed, and the methods enabling the user to interact with the VR learning solution. As of now, there are a variety of methods via which a user can interact with a virtual environment, e.g., mouse and keyboard, hand-held controllers as used in games consoles and multi-touch touchscreen, typically an iPad or tablet. Platforms such as the Oculus Quest and the Microsoft Hololens2 offer the user the ability to engage with the virtual world in 3D at a human scale, allowing the user to control movement within this world through gestures and it is expected that the advent of Apple and Google XR glasses will render this commonplace. However, the prevalence of PC, Mac and tablet devices dictate that a solution must also offer a means to engage with a virtual world via a 2D interface, i.e., the screen. Currently, building virtual learning applications requires gathering material from a wide range of formats, for example documents (such as laboratory manuals), interviews with experts and photographs of the equipment and environment. The virtual solution is then built after manually extracting the information from this material and transcribing it into 3D models of equipment, 3D models of the environment, data sets and electronic training manuals. Previous work at Bloc Digital has shown that creating complex learning scenarios via existing methods can take over a year.  A project considering the maintenance and configuration of a gas-turbine took 3,000 hours to develop and included more than 500 individual training steps. The project required four full time 3D artists/modellers, 3 full time software developers, plus a project manager and a sales manager (personal communication, Bloc Digital, September 2021). Software development was dependent on the existence of equipment models. Hence a bottleneck existed while developers waited for models to be created.

Any updates or corrections require the software to be recoded, issued as a new application and redeployed to the end user. This takes a considerable amount of developer time and is therefore expensive. Due to the number of laboratory simulations required there is a need for lecturers to create their own content for their laboratory tutorials. However, lecturers are not software developers and lack the time and resources to develop virtual applications. Childs et al. (2021) note how a chief inhibitor of widespread adoption by the educational community of AR/VR technologies is the lack of customizable content, mentioning that teachers frequently reported dissatisfaction with the lack of ability to customize educational content for a lesson, both before and during the lesson. Marks and Thomas (2021) also found when evaluating a purpose-built VR laboratory at the University of Sydney that "to enhance innovation up-take it is important to provide in-house content creation support for educators".

Research is currently being undertaken into developing an integrated platform for the semi-automatic creation of digital twins for virtual learning. This paper is an extension of the work described in Palmer et al. (2021) and describes research into the development of an industrial application.  A modelling approach to capture laboratory learning scenarios is considered. Equipment behaviour and procedural information can be modelled generically. Although each individual learning scenario is unique, employing generic models enables information to be reused across scenarios. Initial research into developing electronic training based on an underlying model has been conducted by Torres et al. (2014) who demonstrated the feasibility of the approach.

The objectives of this paper are to demonstrate that:

- Generic models can enable the semi-automatic creation of a virtual learning tools.
- The use of these generic models enables speedier creation of virtual learning tools.
- Digital twins can be applied to Virtual Learning environments.
- A generic modelling approach can be used to implement a digital twin.

This paper is organized as follows. Section 2 provides a literature review. A Digital Twin architecture is presented in Section 3, with Section 4 describing how the architecture is employed to create a VR learning application. A case study demonstrating the approach is described in Section 5. Conclusions are drawn and ideas for future developments are discussed.

## 2. Literature Review

A brief overview of the application areas of VR learning, examples of some commercial application areas of VR training and evaluations of the use of VR for educational purposes are given below. Sub-Section 2.2 defines a digital twin, presents a few relevant examples of digital twins, and explains the need for applying generic modelling to create digital twins.

### *2.1 Virtual Learning Environments*

Joseph et al. (2020) present a list of research articles related to Virtual Reality and its major applications in the education sector. They find that VR has been applied within the following educational sectors: Medicine (Tang et al., 2020), Technology learning, History, Architecture and Natural Sciences. A review of AR/VR in medical practice and education found that the use was primarily in the fields surgery and anatomy for doctors, medical students and interns (Tang et al., 2022). When a VR laboratory was created at the University of Sydney it was found that in the first 2.5 years of use the laboratory was used most by the Faculty of Engineering (53%), followed by the Faculty of Arts & Social Science (23.8%) and Faculty of Science (23.2%) (Marks and Thomas, 2021). In addition, VR learning has also been applied within the disciplines of Architecture (Maghool et al., 2018), Astronomy (Hussein and Nätterdal, 2015), Chemical Engineering (Bell and Fogler, 2004), Computer Assembly (Zhou et al., 2018) and Physics (Yang and Wu, 2010). The University of Nottingham has created a course to learn about computer simulation and VR taught entirely within a VR environment (Virtual Reality World Tech, 2020). Vergara et al. (2017) propose a guide for designing VR learning environments in engineering, emphasizing the need for adequate realism. A list of the top 10 educational virtual reality applications based on user's rating is presented by Babiuch and Foltynek (2019).

A notable example from within the engineering sector is the work of Gupta et al. (2008) who discuss the development of the "Virtual Training Studio (VTS)" which enables a supervisor to create training instructions and trainees to learn assembly operations within a virtual environment. Trainees can learn to recognize parts, remember assembly sequences and correctly position the parts during assembly operations. The VTS consists of three modules: the "Virtual Author" which assists the supervisor in tutorial building; the "Virtual Workspace" which loads and executes the tutorials created in the "Virtual Author" within the virtual environment; and the "Virtual Mentor" which monitors the trainees within the virtual environment, providing them with adaptive hints and error messages. The "Virtual Author" generates text instructions by analyzing the supervisor's movement, object collisions and feature alignment.  The VTS was used to develop tutorials on the assembly of a small military rocket and a model airplane engine (Brough et al., 2007).

Commercial applications of VR training are beginning to come into wider use, a trend accelerated by the pandemic.  Pixo (n.d.) claim to offer the world's largest collection of off-the-shelf virtual reality training.  Courses are provided in the subject areas of workplace safety, manufacturing, energy  and construction, first aid soft skills and meditation.  A manufacturing training example describing how maintain an industrial water reclamation system is shown (Pixo 2017, video).

As with academic use, commercial applications of VR training seem to be mostly employed within the engineering sector although some examples within the service industries are beginning to be developed.  In the manufacturing engineering domain Rolls-Royce supplies an instructor-led immersive live Virtual Training tool to train customers how to service its BR725 engine and undertake non-routine maintenance. This tool is not intended to completely replace practical training but to provide greater flexibility and the eliminate the need to ship a full-size training engine (Rolls-Royce, n.d.).   Qatar Airways Company Q.C.S.C. (Qatar Airways) was the first operator to participate in the program (Kucinski, 2019). GlaxoSmithKline use VR to provide maintenance training of their reactors (Bloc Digital, 2017).   In the chemical engineering domain BP and ExonMobil user VR for safety training (Viar360, n.d.).

Within the service industries Hilton Hotels use VR training as a cost-effective method to provide their employees with realistic experiences (MetaQuest, 2020). Walmart is also planning to use VR for scenario training. Kentucky Fried Chicken is planning to use VR as training in food preparation and UPS will begin training student delivery drivers using virtual reality headsets (Viar360, n.d.). Examples of VR used in corporate training (created by 360° VR) are: robbery emergency training; Diversity and Inclusion training, soft skills development for employee coaching and employee onboarding (Roundtable Learning, n.d.).

Marks and Thomas (2021) found that when a purpose-built VR laboratory was introduced at the University of Sydney 71.5% of students surveyed reported enhanced learning outcomes. They found that disadvantages reported were: headaches, dizziness, blurred vision, the weight of the head-set and that it did not fit over glasses. Hussein and Nätterdal (2015) reported that a benefit of VR was that it allowed users to experience scale (of planets in an astronomy application) but they also found that a minority of students experienced motion sickness and minor headaches. Zhou et al. (2018) found no different in learning completion time between a control group who used real world equipment and VR learners. Nesenbergs et al. (2021) conducted a review to determine if VR/AR technologies are beneficial to performance and engagement in remote learning, finding mostly positive results.

It can be seen that VR is widely applied with education and is mostly positively received by students. Problems with motion sickness can be overcome by the use of appropriate software refresh rates to reduce display flickering (Chang et al., 2020). Given the advantages of VR described in the introduction, there is a need to develop more effective method of content creation for learning applications.

### *2.2 Digital Twin*

A digital twin can be defined as a digital or virtual representation of a real or potential product, system, process or value chain, although how the term is applied throughout the product lifecycle may vary. However, its principal use is to foster a common understanding of the system in question. Digital twins can be applied in many domains e.g., manufacturing, industrial Internet of Things, Healthcare, Smart Cities, Automobile and retail (Augustine, 2020). An overview of digital twins, their history, advantages, markets and applications is given by IBM (2022). A systematic literature review is presented by Jones et al. (2020) who define 13 characteristics as follows: “Physical Entity/Twin; Virtual Entity/Twin; Physical Environment; Virtual Environment; State; Realisation; Metrology; Twinning; Twinning Rate; Physical-to-Virtual Connection/Twinning; Virtual-to-Physical Connection/Twinning; Physical Processes; and Virtual Processes”. The "Physical Entity/Twin" exists in the "real world" e.g., a vehicle, part, product, system. The "Virtual Entity/Twin" is computer generated representation of the "Physical

Entity/Twin". The "Physical Environment" denotes the "real-world", whilst the Virtual Environment comprises the digital domain within which the "Virtual Entity/Twin" exists. The physical/virtual/entity/twin possesses measured parameters values which define "State". "Metrology" is the act of measuring "State". "Realisation" is the act of changes the "State" value. Twinning is the act of synchronising the virtual and physical states. The "Twinning Rate" the frequency at which twinning occurs, is only described in the literature as being in "real-time". "Physical-to-Virtual Connection/Twinning" is the means by which the state of the physical entity is transferred to the virtual environment e.g., Internet-of-Things sensors, web-services. "Virtual-to-Physical Connection/Twinning" is the flow of information and processes from the virtual to the physical e.g., display terminals and PLC's. "Physical processes" are processes with which the "Physical Entity/Twin" interacts. "Virtual processes" are processes with which the "Virtual Entity/Twin" interacts.

No applications of digital twins within VR learning environments have been found within the literature. However, Kaarlela et al. (2020), describe three use cases of VR and digital twins for safety training. Kuts et al. (2018) present a “Teaching factory” to train operatives. Vahdatikhaki et al. (2017) discuss a VR equipment training simulator. Although the authors do not recognise this application as a digital twin, sensor data from construction site equipment is used to generate a VR scene.

The application of the digital twin approach within a VR learning environment is limited by the complexity of creating equipment models which determines cost effectiveness. The advantages of applying a modelling approach to implement the digital twin is that the generation of equipment and operating procedures is facilitated through semi-automated creation based on generic models. The use of generic models aids re-use and reduces development costs. Generic model use removes the need for repetitious low-level coding, preventing duplication and the inconsistencies which can result from manual coding.

## 3. The Digital Twin Platform

The Digital Twin platform consists of three main elements:

1) The Digital Twin Builder application which enables a lecturer to specify laboratory tutorials via an easy to use drag and drop interface and store these specifications in a serialised file format for re-use.

2) Three processing pipelines (data, geometry, and process) which characterise incoming information and make it available within the drag and drop interface.  More information on each of the pipelines is provided in the subsections below.

3) The Digital Twin Player, which provides the VR learning application,

enabling students to interact directly with the digital twin produced and to share results with fellow learners and the lecturer. This collaboration process is shown in Figure 1.

*Figure 1 about here*

The scenario definition file created with the Digital Twin Builder separates the virtual world creation from the virtual world viewer. This allows the same digital twin configuration to support multiple form factors and hardware platforms, and engage users in an optimal manner (e.g., XR glasses, tablet etc.).

The architecture of the Digital Twin platform is shown in Figure 2. The figure shows the process of importing data, geometry and processes into the Digital Twin Builder and producing Digital Twin specifications which can be used with the Digital Twin Player(s) to view, interact with, and collaborate around.

*Figure 2 about here*

### *3.1 The Data Pipeline*

With any Digital Twin it is essential that it can behave in a realistic manner based on external data events or stimuli. The Digital Twin Builder processes, parses and transforms the data into a suitable generic model format to enable the updating of object behaviour in the Digital Twin Player. The Data Pipeline has the ability to execute incoming data flows from multiple sources. Within the Digital Twin

Builder the lecturer links the data models imported by the Data Pipeline to relevant laboratory equipment objects. Data is imported within a spreadsheet and the lecturer selects columns of interest from a dropdown menu within the Digital Twin Builder. A variable number of columns may be selected.  This section describes the data types employed by the Digital Twin Platform, the Digital Twin Builder data services for processing the data types and how the Digital Player displays the data types.

The VR learning application utilises data to produce simulation results. The possible range of data comprises time series data, tabular data, sensor data, service data, and simulation data. Common data types and their details are given in Table 1.

*Table 1 about here*

The following paragraph provides explanations for the terms given in Table 1 where clarification is needed.  Explanations for the data types in Table 1 are: csv, comma separated file; CAD, Computer-Aided Design for three-dimensional modelling; IoT, Internet of Things sensor data; Mesh, a 3D description of an object. A detailed description of the difference between CAD and Mesh data types is given in section 3.2.1.   An object texturing display consists of an array of colour pixels which gives more realist appearance of 3D when applied to VR objects. Definitions for the examples of data types in Table 1 are:

- “Response Surface”, where several input variables to a model or simulation are plotted to show the variation of one or more output variables.
- “Test Article” is the physical object tested.
- “As-Designed Geometry”, is the shape the designer intended.
- “Design for Manufacture Geometry" is the design that is possible to make. It may slightly differ from the As-Designed Geometry to simplify manufacture.
- “As-Made Geometry” is the product of the manufacturing process.

The “Performance, Trend Analysis and Limits” use shows the performance of an item or system over time. The “Review” use consists of presenting data for review purposes e.g., a Critical Design Review.

Currently the Digital Twin Builder contains data services for processing and parsing three types of data as follows:

(1) Time series data stream is parsed with a given time schedule. This enables the task of updating the object behaviour to performed efficiently within a sequence of time slots.

(2) Tabular data is first loaded in a data buffer and then executed tuple by tuple. This technique renders the tabular data execution process more efficient by providing a simplified, memory friendly architecture.

(3) Sensor data is modelled as a data stream with a series of points. The service for sensor data is capable of parsing the data stream quickly and changing the object behaviour in real time.

The prototype is capable of evaluating mathematical expressions using external data streams as terms, e.g., the power generated by a model of a wind-tunnel can be calculated based on a wind-speed measurement.  The data source is associated to a data input channel and the channel data is associated with labels via the dropdown menu in the Digital Twin Builder, e.g., the sensor data from an anemometer may be labelled as sensor_wind_speed. Combining this with constants, such as the average_density_of_air_mass_at_sea_level, expressions can be evaluated to produce:

$$\text{Force_of_Wind} = \text{average_density_of_air_mass_at_sea_level}\ \ \text{sensor_wind_speed}^2$$

This is stored as a data-calculation output and the behaviour of an equipment object can be made to react to the output, for example, by causing a kite to fly.

The data source locations are stored in a “Data Lake” which is used by the Digital Twin Player to load the data (see Figure 3 below).

*Figure 3 about here*

The Data Pipeline has a visualization module which enables data to be displayed within the Digital Twin Player as graphs, textual displays and as desired equipment behaviours. In order to display information for the user to explore, the Digital Twin Player provides four customizable data display types:

(1) The Text display type is used to display a static text, e.g., label, name, value for a specific object.
(2) The Dynamic Text display type shows the variable data attached to an object, e.g., position, value.
(3) The Image display type is used to visualize image data, e.g., a circuit design, data flow design.
(4) The Graph Display type comprises multiple graph types for different types of data. The relevant type of graph which displays with respect to the data type is shown in Table 1.

### *3.2. The Geometry Pipeline*

Within a virtual lab environment or Digital Twin, environments and equipment are represented using geometry objects. In order to render this geometry performantly across the range of hardware platforms supported, the Digital Twin Builder must use a data format suitable for real-time rendering. This section commences by describing the geometry modelling technique (Polygon Mesh Creation) used by Digital Twin Builder and how these models were traditionally created. The new automated production technique of these models by the geometry pipeline is introduced (3.2.2). The machine learning reduction process employed by the production technique which reduces the final model size output is described in 3.2.3. In 3.2.4 collection of training data for the machine learning process is discussed). This section concludes by justifying the machine learning approach (Random Forest) employed by the Digital Twin Builder (3.2.5).

### *3.2.1 Polygon Mesh Creation*

Digital Twin Builder utilizes polygonal meshes to enable real time rendering. Polygonal meshes are a lightweight means of representing arbitrary 3D geometry, and are commonly used in real-time applications such as 3D video games and animations (Akenine-Moller, 2018). A polygon mesh approximates an object as a finite set of vertices and polygons (triangles). A greater number of polygons produces a more accurate approximation of the real object at the expense of increasing the computational burden required to render that approximation. As such, the creation of meshes for use in a digital twin is often a trade off between visual accuracy and performance considerations. In traditional digital twin production, geometry is created by skilled 3D artists. This is one of the largest expenses in the production process. In order to reduce these costs and open up digital twins for use in teaching environments, an economical means of creating accurate polygonal meshes is required.

While polygonal meshes are common in the gaming and animation industries, applications in research and education are more likely to use and have access to computer aided design (CAD) formats for geometry. CAD models use spline curves to produce perfectly accurate geometry, however the algorithms involved in rendering CAD precludes the use of all but the simplest models in real-time applications. Importing CAD models into digital twins thus involves the conversion of CAD data to a polygonal format. Such a conversion is often performed manually in order to balance accuracy with performance. While semi-automated tools for CAD to mesh conversion exist, many still require a level of knowledge of 3D modelling which users in technical and education sectors are unlikely to possess.

### *3.2.2 Automated Production*

To allow users in education to create suitable geometry for virtual lab environments, Digital Twin Builder contains a *geometry pipeline* to automate the production of visually accurate, low polygon count meshes. This pipeline accepts geometry from several CAD and polygonal mesh formats, converts these to the chosen FBX format, and then uses a machine learning based approach to selectively reduce the detail of certain aspects of the geometry to reduce the total polygon count without unduly affecting the visual quality of the overall model. Where the model consists of multiple parts, each part is reduced independently, and parallelization is employed to reduce the overall runtime of the process. In cases where parts are duplicated (i.e. in which several parts have identical meshes but different transforms), only one part (the original) is reduced. The original is then duplicated and the duplicates transformed as necessary. This duplication process both reduces the runtime of the pipeline and the size of the finished model, as duplicate parts are simply stored as shallow copies of the original. Finally, the original and duplicated parts are rebuilt to produce the low polygon model. Figure 4 shows the overall process employed by the pipeline, while Figure 5 highlights the machine learning based reduction process.

*Figure 4 about here*

*Figure 5 about here*

The geometry pipeline is predominantly written in Python 3.5 and uses two commercial off-the-shelf software packages to deal directly with the 3D objects. Autodesk 3DSMax (Autodesk, 2021) is used to perform the initial conversion from CAD to high poly mesh (the "Convert" step in Figure 4). This is a simple format conversion process, with 3DSMax running as a Python subprocess according to a script. MooTools Polygon Cruncher (Mootools, 2021) is used to convert the high poly mesh to the low poly mesh. Polygon Cruncher uses the method of progressive mesh decimation (Hoppe, 1996) to reduce mesh polygon counts, and exposes a number of parameters to control this process for a given mesh. The machine learning framework shown in Figure 5 is used to predict which set of parameters will give the best results for a given mesh, and verify these results after conversion is complete.

*3.2.3 Details of the Machine Learning Technique employed*

To automatically produce visually accurate low polygon meshes requires the geometry pipeline to assess the (subjective) visual quality of a mesh based entirely on (objective) measurements of mesh topology. This is achieved using a framework built around three machine learning models, as shown in Figure 5. For a given high poly mesh, this framework first extracts several *shape metrics* to codify the mesh topology as a set of floating-point values. These metrics are then merged with every possible configuration of Polygon Cruncher parameters to create a data set in which every observation represents a unique combination of topology and settings. These data are then passed to a pair of random forest regressors to predict the visual quality and polygon count of the models resulting from every combination of Polygon Cruncher parameters. These models predict a quality score and polygon ratio respectively. Both values are normalized from 0 to 1, and a final score is found as the ratio of quality score to polygon ratio. The data set is then sorted in decreasing order of this score. The highest scoring set of parameters is then applied within Polygon Cruncher to produce the *trial* low poly mesh. A second set of shape metrics is then extracted for this mesh. In addition, a set of *shape ratios* is calculated between the high and low poly shape data sets, and a set of *similarity metrics* are calculated based on the topology of both the high and low poly meshes. All of the metrics used in the model are listed in Table 2. The three sets of metrics (high poly shape, low poly shape, similarity) are then fed to a random forest classifier which determines if the low poly mesh is sufficiently representative of the high poly mesh. If this test passes, the mesh is accepted for the final model. If not, the next highest scoring set of Polygon Cruncher parameters is used. This process repeats until all parts have passed the test, or until any parts have failed 10 times, at which point we assume the object cannot be reduced and resort to using the high poly mesh for that part.

*Table 2 about here*

### *3.2.4 Collection of Training Data*

Training data for the three machine learning models were collected as follows: First, a representative set of 37 mechanical objects was produced. Next, every object was processed through Polygon Cruncher using every potential set of parameter values, giving 108 meshes per object and 3,996 meshes in total. The polygon counts of all 3,996 meshes were collected and used to train the polygon prediction model. For quality prediction, the models were trained against human judgments. Judgments were collected using a companion application. Each of the 3,996 meshes was duplicated 5 times to give 19,980 training meshes. Each training mesh was displayed in the application alongside the original (high poly) mesh it was produced from, as shown in Figure 6. These training meshes were then distributed to a group of 23 volunteers, who ranked each trial mesh with a score of “ruined”, “bad”, “good” or “perfect”. These scores were later converted to numeric form and normalized to give the quality score for the mesh. The training data used for quality prediction were also gated for use in the similarity checking model, where results of “good” or “perfect” result in a mesh being accepted. The reuse of this data set in both models is justified given that the similarity checking model has a considerably greater number of predictor variables than the quality prediction model.

*Figure 6 about here*

*3.2.5 Reasons for choosing the Random Forest Approach*

As mentioned above, the geometry pipeline relies on three integrated random forest models. Random forests were chosen for four main reasons: Firstly, these methods are more explainable than deep learning methodologies, allowing for a better understanding of the decisions they make regarding mesh quality. Second, random forests are generally seen to be less susceptible to overfitting than many machine learning methods. Third, random forests are capable of reasonable accuracy on relatively small data sets. Given the difficulty in collecting suitable training data in large quantities, this was seen as a major advantage in this application. Finally, given the role of the similarity checking model as a guarantor of the quality prediction model's accuracy, random forests were seen as more than accurate enough to reliably produce visually accurate low polygon meshes. Combining the quality prediction and similarity checking models gives a 97% acceptance rate for low polygon meshes.

Using the machine learning framework shown in Figure 4, coupled with the parallelization approach shown in Figure 5, the geometry pipeline is capable of producing visually accurate low polygon count meshes of large, complex equipment an order of magnitude faster than this can be achieved using traditional manual approaches. For example, the model shown in Figure 7 consists of 106 parts (of which 39 are unique). To model this manually would take approximately 8 hours. Running on a dedicated server with 24 Intel Xeon 2.20GHz CPU cores and 64GB of RAM takes approximately 23 minutes.

*Figure 7 about here*

### *3.3 The Process Pipeline*

A VR learning application will contain processes to guide a student as to which tasks are needed to undertake the required assignment or tutorial. A process is defined as an ordered or unordered set of steps. Ordered steps contain a link to the next step in the process. In order for a training procedure to interact with a Digital Twin it must be formally modelled. The training procedure directs actions to change the state of the Digital Twin. As the user executes the training procedure the interaction states of the Digital Twin equipment items are updated. Manual creation of these processes is time-consuming and error prone, requiring translation of tutorial instructions into machine readable data and linking this data to equipment objects, user inputs, and physical processes.

The Process Pipeline captures procedure information in domain specific models generated from human readable structured forms. The domain specific models enable hierarchical ordering of process stages, inclusion of equipment references, and multiple forms of interdependency between process stages and equipment states. This approach enables the procedures to be directly defined by a lecturer without the need for requirements gathering, coding and re-iterating until a satisfactory application is achieved.

Two types of process model step exist: procedure and instruction. Procedures contain steps (i.e., a procedure can contain procedures, instructions, or both procedures and instructions). A procedure contains an identifier and a description of its purpose e.g., “(1) Assemble Electrical Lab”.

An instruction directs an action to change the state of the Digital Twin. An instruction refers to one or more equipment items within the Digital Twin. For example, an instruction can act upon on item, e.g., “turn dial to 'on' position”, connect two items together, e.g., “use a cable to connect DC motor to ammeter” and position three items, e.g., “Place Voltmeter between cable Connect5 and cable

Connect4". Instructions contain a condition which must be satisfied in order to progress to the next step. These conditions are based on the user altering an equipment item to a defined state (such as pressing a button to "on"), external factors (wait ten seconds) or calculation results which can be retrieved from a data source. To achieve a state change of the Digital Twin the instruction references one or more scenario object's interaction state.

An instruction contains:

- An identifier;
- A description of its purpose, e.g. "Use a cable to connect DC motor +ve to 120V source -ve";
- A list of equipment objects which the instruction refers to. The equipment objects within the list are named as ActionObject, TargetObject and TargetObject2;
- The interactions of the equipment objects which the instruction monitors;
- A state which indicates when the instruction is complete;

The user undertakes the action indicated by the instruction within the Virtual Training application, changing the states of the equipment items interactions to the instruction state. Ordered procedures and instructions contain an identifier of the next step in the sequence. An overview of the Process Pipeline model is given in Figure 8 below.

*Figure 8 about here*

There is an assumption within the process pipeline that a virtual environment model (i.e., the laboratory) exists. The process pipeline contains a “Scenario checker” which ensures that the process steps do not have any missing information and refer to equipment item with the Digital Twin.

## 4. Applying the Digital Twin Architecture to create a VR Learning Application

To create a VR learning application using the Digital Twin architecture the following information is needed: a list of equipment and 3-D representations of these equipment items for display in the virtual world; a virtual environment in which to place the equipment (e.g. workbench, laboratory area); the location of the equipment within the virtual world; the location of VR equipment interactions (i.e. the area of the equipment which interfaces with another item during assembly or with the user during operation) and user interaction type (i.e. whether to push a button, turn a dial, add a cable); a list of tutorial instructions; and data to enable simulation.

The Geometry Pipeline provides 3-D equipment model representations for the equipment items and virtual environment in the form of FBX files (Autodesk, 2021). The hierarchy of the FBX file is defined as follows: At the root level, the file contains one or more *objects*. Each object may either contain other objects or *meshes (*but not both). A mesh is a three-dimensional, polygonal representation of a piece of geometry, made up of sets of vertices and faces. Meshes are always leaf nodes within an FBX. As well as potentially containing meshes and other objects, each object contains a *transform.* This is a set of vectors representing the position, rotation and scale of the object relative to its parent.

The 3-D equipment model representations are imported into the Digital Twin Builder as reusable generic object models.  The FBX hierarchy is reproduced as follows:

(5) Every FBX file is referred to as a *model,* which now contains a *transform*

(6) The root of all models is the *scenario*

(7) The *objects* within an FBX file are referred to as *parts*

(8) As well as meshes and other parts, models and parts may contain *interactions.* Interactions contain transforms, are always leaf nodes, and are defined further below.

Figures 9 and 10 show the two hierarchies.

*Figure 9 about here*

*Figure 10 about here*

Interactions are the components of the scenario which allow models and parts to respond to each other and to the actions of the user. As shown in Figure 10, each interaction belongs to a model or part, and a given model or part can contain any number of interactions. As well as a transform, which defines the relative position, orientation and scale of the interaction to its parent, and interaction also defines a *type* and one or more *parameters*. The *type* of an interaction defines how it responds to other interactions or to the user. Many interaction types exist, including user-based interactions such as buttons, dials and snap connectors, data driven interactions such as text and image displays, and model behavior defined interactions such as rotation components. The parameters of an interaction depend on its type. For example, a rotation component has parameters representing the axis of rotation and the speed with which the object rotates around this axis. A dial interaction has a parameter representing each of the states to which the user may set the dial. Finally, interactions may also contain data processors, allowing the interaction to communicate with other data sources, data processors, and data displays within the scenario.

The Digital Twin Builder provides a user-friendly form and a drag and drop interface to enable VR equipment interactions to be defined for the generic objects. To create a VR learning application a scenario is created within the Digital Twin Builder and a virtual environment is added to it. Object models are instantiated into the scenario and location information is added to these equipment instances via a form and drag and drop interface. The Data Pipeline and Process Pipeline enable simulation data and tutorial instructions to be added to the VR learning application.

The Digital Twin Builder stores the scenario created within a definition file (see Figure 11) which contains links to the equipment object FBX files to allow the software to render the geometry. In addition, the definition file also reproduces object hierarchy of the FBX file in a manner which allows for modifications to be made to the object within the digital twin without modifying the FBX file itself. This allows the same FBX file (i.e., the same geometry) to be used in a different manner across multiple digital twins.

*Figure 11 about here*

## 5. Results of a Case Study: An Electrical Laboratory Tutorial

The brief shown in Figure 12 was provided as a test case for the Digital Twin Platform.

*Figure 12 about here*

The Unity development platform (Unity 2019.3) was used to implement the Digital Twin platform. Figure 13 shows a screen shot of the electrical laboratory tutorial being defined within the Digital Twin Builder. Lists of the equipment instances and procedures which have been created to fulfil the case study brief are displayed within the left-hand panel of the Digital Twin Builder. The virtual learning scenario created is shown in the centre panel. The arrows protruding from the equipment instances indicate VR interactions. The right-hand panel of the Digital Twin Builder shows the values of a dial interaction which has been selected. The right-hand panel provides an example of the forms via which equipment and process information can be specified and edited. The centre panel shows a sphere, constituting the drag and drop interface providing an alternative means to provide equipment location, size and rotation information.

*Figure 13 about here*

Figure 14 displays the electrical laboratory within the Digital Twin Player which is shown running on a P.C.  It can be seen that the panel holding the instructions on the left-hand side of the Player colour upon completion of the directed action.

*Figure 14 about here*

Figure 15 presents a case study equipment item within the Digital Twin Player.  The Digital Twin Player shows the DC motor with its data simulation results displayed as a graph. As the Digital Twin Player runs the DC motor rotates simultaneously in response to the values of the speed data displayed.

*Figure 15 about here*

The following feedback was received from the lecturer conducting the virtual electrical laboratory tutorial. "Using the virtual learning environment, the setup time is less compared to that of the actual experiment, while assessing the same learning outcomes as the actual system (face-to-face within the laboratory). In the actual experiment, some time is required (approximately 5 minutes) to correctly connect the Electrical machine, as provided in the instruction manual. This takes about a minute for the virtual case. Usually, due to bearing friction, wear and tear of rotational coupling joints and general uncertainties, the results obtained from the actual experiment slightly differ from those of the virtual experiment. This is totally expected, though. The main benefit of the virtual experiment is the possibility of it being conducted anywhere, unlike the actual experiment, which needs to be done on campus. The virtual experiment has been especially important during the COVID-19 pandemic since it allowed learners to experience something close to the physical lab and meet the same learning outcomes as the actual experiment."

This case study demonstrates that it is possible to create a Virtual learning environment based on a Digital Twin. Exact timing measurements have not been possible for this prototype version as the case study formed part of a "test and develop lifecycle" for the Digital Twin architecture, so underwent several revisions. However, the author who created the case study estimates that the use of generic models to enable the semi-automatic creation of a virtual learning tool reduces the development time by a factor of 12 from weeks to hours.

## 6. Conclusions

An architecture for a Digital Twin platform has been defined which enables a virtual learning application to be semi-automatically created. The Digital Twin Architecture utilizes a generic modelling approach which allows information re-use between virtual tutorials, thus saving work for lecturers. The current version utilizes current equipment data. It is envisioned that a future version will allow students to control equipment remotely. It was found that whilst this new approach can reduce the time taken to create a VR learning application from weeks to hours, the most time-consuming feature is the need to accurately position the VR interactions to enable equipment items to connect together, necessitating iteration between the Digital Twin Editor and Player to check location.

## 7. Future Work

Future work to enhance the functionality of the Digital Twin includes:

(1) Research into the best form of interface to create the VR interactions and into whether semi-automatic positioning is possible.

(2) Investigating a cloud-based service for the Digital Twin for remote access (Hasham et al.,2011; Baker et al., 2012, Van Lingen et al., 2005),

(3) Edge enhanced Digital Twins for performance and real time analytics (Yaseen, et al., 2018; Ali et al., 2020; Liu et al., 2020; Kiani et al., 2020, Hill et al., 2017).

(4) APIs to connect the Digital Twin with simulation software such as LabView (National Instruments, 2021), NASTRAN (Autodesk, 2022).

(5) Further development of the process pipeline.

(6) The use of advanced object recognition and contextual understanding.

(7) Research into human factors and user experience.

Development of the process pipeline will focus on three areas: Firstly, the language definition will be extended to allow for more complex process structures. This will enable steps to be combined using logical (AND, OR, XOR) expressions. Secondly, research will be undertaken to allow for direct conversion of less structured process data (e.g., maintenance logs written by service engineers) into Digital Twin processes. This conversion will require the consideration of natural language processing and domain specific modelling techniques. The third area will consider facilitating procedure capture from a user. The user will be recorded interacting with a virtual scenario in a "preview version" of the Digital Twin Player and instructions will be created from the user's actions. This will reduce the time it takes to input procedures and allow tacit procedural knowledge to be captured. When considering an activity it is easy to overlook an action, the need for which becomes immediately apparent in real (or virtual) life. Some initial work in this area has been developed by Gupta et al. (2008) who generate text instructions by analyzing an expert's actions. Consideration will need to be given as to how the simple actions captured can be built up into hierarchical procedures or how alternative or parallel instruction pathways can be captured.

The object recognition work will use machine learning models to consider how to identify and pre-populate likely interaction points on imported assets. Parts imported will be identified (e.g., "this is a gear"). The context of each part relative to its surroundings will be derived (e.g., "gear A is meshing with that gear B") and parameters calculated based on this ("the reduction ratio between these gears is 3:2). Relevant interaction points will be pre-populated. For example, in the case of gear A, a rotation interaction will be added with a data link that showing that gear A rotates at 1.5 times the rate of gear B (Rucoo et al., 2019). This approach will reduce the time it takes for a user to specify equipment information.

The human factors research will consider methods of presenting Digital Twin information on screen, in AR and in VR, which enable optimal use of the user's and equipment time, user collaboration and adaptive training. In future more importance will be given to developing interfaces for Mobile Devices/Tablets as students prefer using these devices to PC's. There will be an emphasis on developing an intuitive holographic interface for accessing information without being intrusive and suitable for industrial scenarios.

There is a need for users to collaborate within a shared virtual workspace, in the same way as in the real world. For example, one student might position and steady a piece of laboratory equipment whilst another tightens the equipment fixings. Adaptive training could allow a user to perform a procedure within the Digital Twin Player. A facility could be provided within the player to compare the user's actions with a procedure provided by an expert and the user could be informed about any possible deviations from the expert's procedure. Hints could be provided to enable the user to perform the procedure. Textual pop ups could also be provided to explain the required VR interactions for student who lack experience in using graphical interfaces. The user's actions could also be checked against safety requirements. This would enable a user to learn from mistakes, but within a safe virtual environment. Trial and error has been shown to improve memory (Hays, Kornell and Bjork,2013; Cyr and Anderson, 2018).

## Acknowledgements

This work was supported by Innovate UK Knowledge Transfer Partnerships and Bloc Digital under Grant agreements no.11936 and 11251.

## Declaration of Interest Statement

This research is sponsored by Bloc Digital and may lead to the development of products. I have disclosed those interests fully to Taylor & Francis, and have in place an approved plan for managing any potential conflicts arising from this arrangement.

**Word Count**: 8900 words

| Data Source | Data Type | Displays | Examples | Uses |
|---|---|---|---|---|
| Design Data | csv | Graphs/Bar Charts | Simulation Data | Performance, Trend Analysis and Limits |
| | 2D Data | 3D plots | Response Surface | Performance, Trend Analysis and Limits |
| | CAD | 3D Object | As-Designed Geometry | Review |
| Manufacturing Data | csv | Graphs/Bar Charts | Machine tool performance, consumable levels | Performance, Trend Analysis and Limits |
| | IoT | Graphs/Bar Charts | Machine tool performance, consumable levels | Performance, Trend Analysis and Limits - ordering of consumables |
| | Mesh | 3D Objects | Design for Manufacture Geometry | Comparison of as-designed and as-made |
| | Point Clouds | 3D Objects | As-Made Geometry | Comparison of as-designed and as-made, Review |
| Test Data | csv | Graphs/Bar Charts | System performance | Performance, Trend Analysis and Limits |
| | IoT | Object Texturing | Sensor Data | Performance, Trend Analysis and Limits |
| | Mesh | Object Texturing | Test Article | Review |
| Service Data | csv | Graphs/Bar Charts | In-Service performance, i.e. utilisation | Performance, Trend Analysis and Limits - Data |
| | | | rates | Analytics |
| | IoT | Graphs/Bar Charts, Object Texturing | Real time system configuration data | Performance, Trend Analysis and Limits - Data Analytics |
| | Mesh | Object Texturing | Scanned Objects | Review, Performance, Trend Analysis and Limits - Data Analytics |

**Table 1.** Digital Twin Platform Datasources

| Symbol | Metric | Type |
| --- | --- | --- |
| $L_1$ | Bounding box length/width ratio | Shape Metric |
| $L_2$ | Bounding box length/depth ratio | Shape Metric |
| $\varphi$ | Sphericity | Shape Metric |
| $\rho_{Box}$ | Bounding box density (mesh volume / bounding box volume) | Shape Metric |
| E | Shape Efficiency $E = R_M / R_B$ | Shape Metric |
| $A_{skew}$ | Skewness in polygon area | Shape Metric |
| $A_{kurt}$ | Kurtosis in polygon area | Shape Metric |
| $A_{COV}$ | Coefficient of variance in polygon area | Shape Metric |
| $\alpha$ | Mesh connectivity (ratio of polygons to vertices in mesh) | Shape Metric |
| $R_V$ | Low poly / High poly volume ratio | Shape Ratio |
| $R_A$ | Low poly / High poly surface area ratio | Shape Ratio |
| $R_\varphi$ | Low poly / High poly sphericity ratio | Shape Ratio |
| $R_\rho$ | Low poly / High poly bounding box density ratio | Shape Ratio |
| $R_E$ | Low poly / High poly shape efficiency ratio | Shape Ratio |
| $R_\alpha$ | Low poly / High poly mesh connectivity ratio | Shape Ratio |
| $R_{skew}$ | Low poly / High poly polygon area skewness ratio | Shape Ratio |
| $R_{kurt}$ | Low poly / High poly polygon area kurtosis ratio | Shape Ratio |
| $R_{COV}$ | Low poly / High poly polygon area coefficient of variance ratio | Shape Ratio |
| D | Average distance between equivalent points | Similarity Metric |
| $\Delta N$ | Average normal vector deviation | Similarity Metric |
| $\Delta\theta$ | Average dihedral angle deviation | Similarity Metric |

**Table 2**. Shape and similarity metrics used in the geometry pipeline to ensure mesh visual quality. Note that none of these metrics encode the scale of the object, to ensure the method is scale invariant.

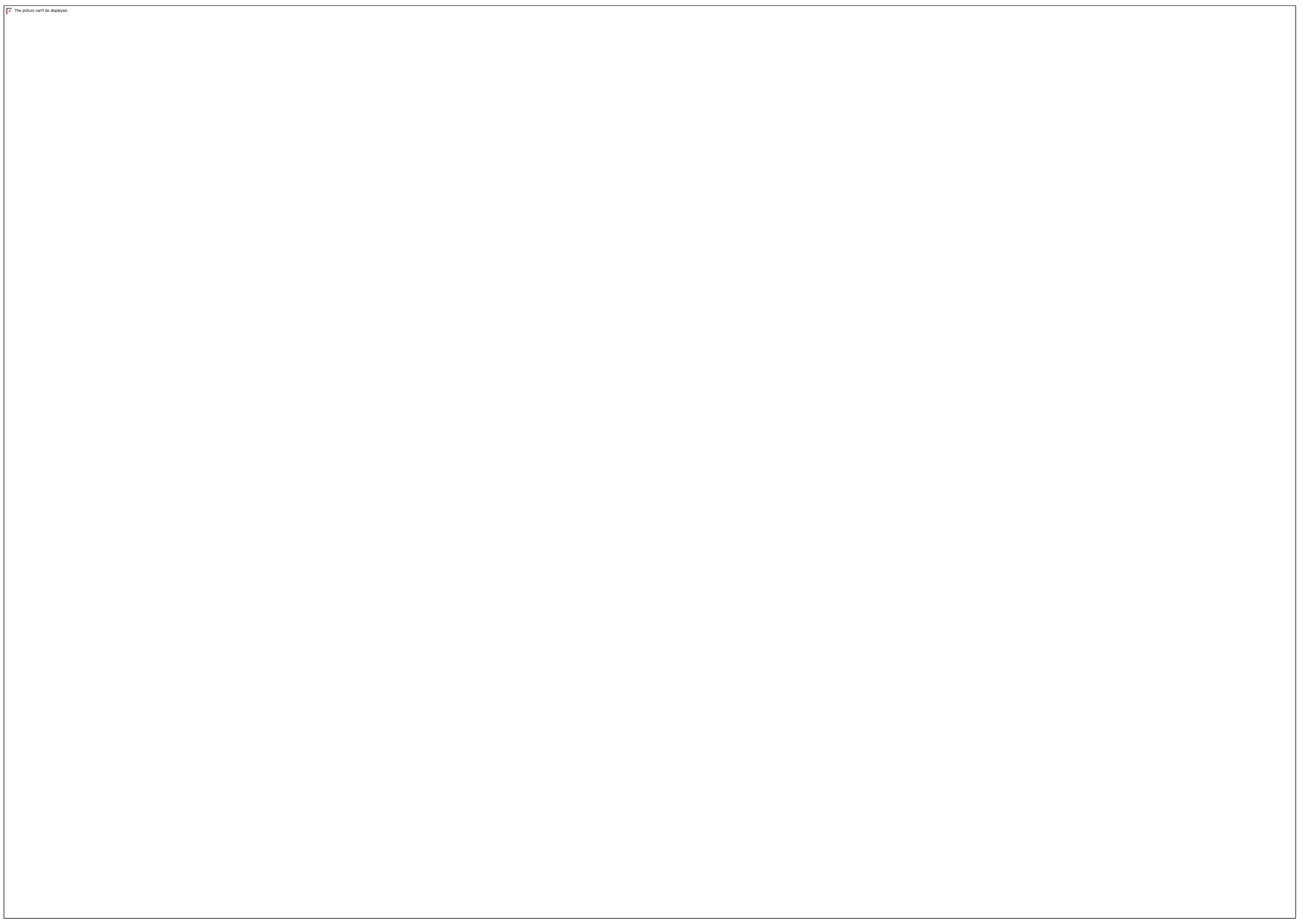

**Figure 1.** Collaboration with the VR Learning Environment.

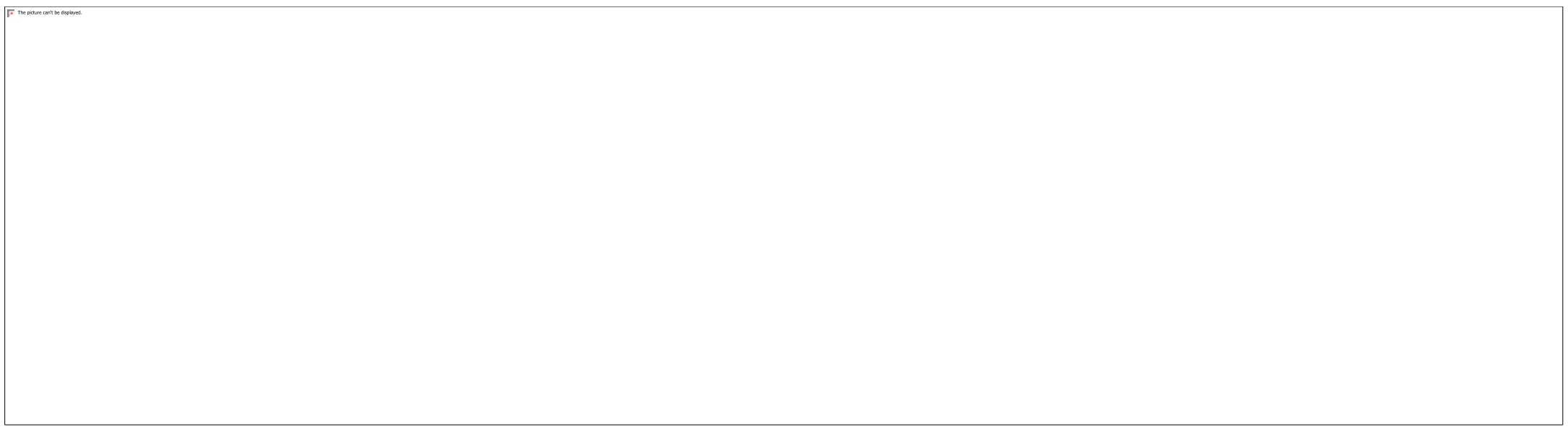

**Figure 2.** Digital Twin Builder Architecture.

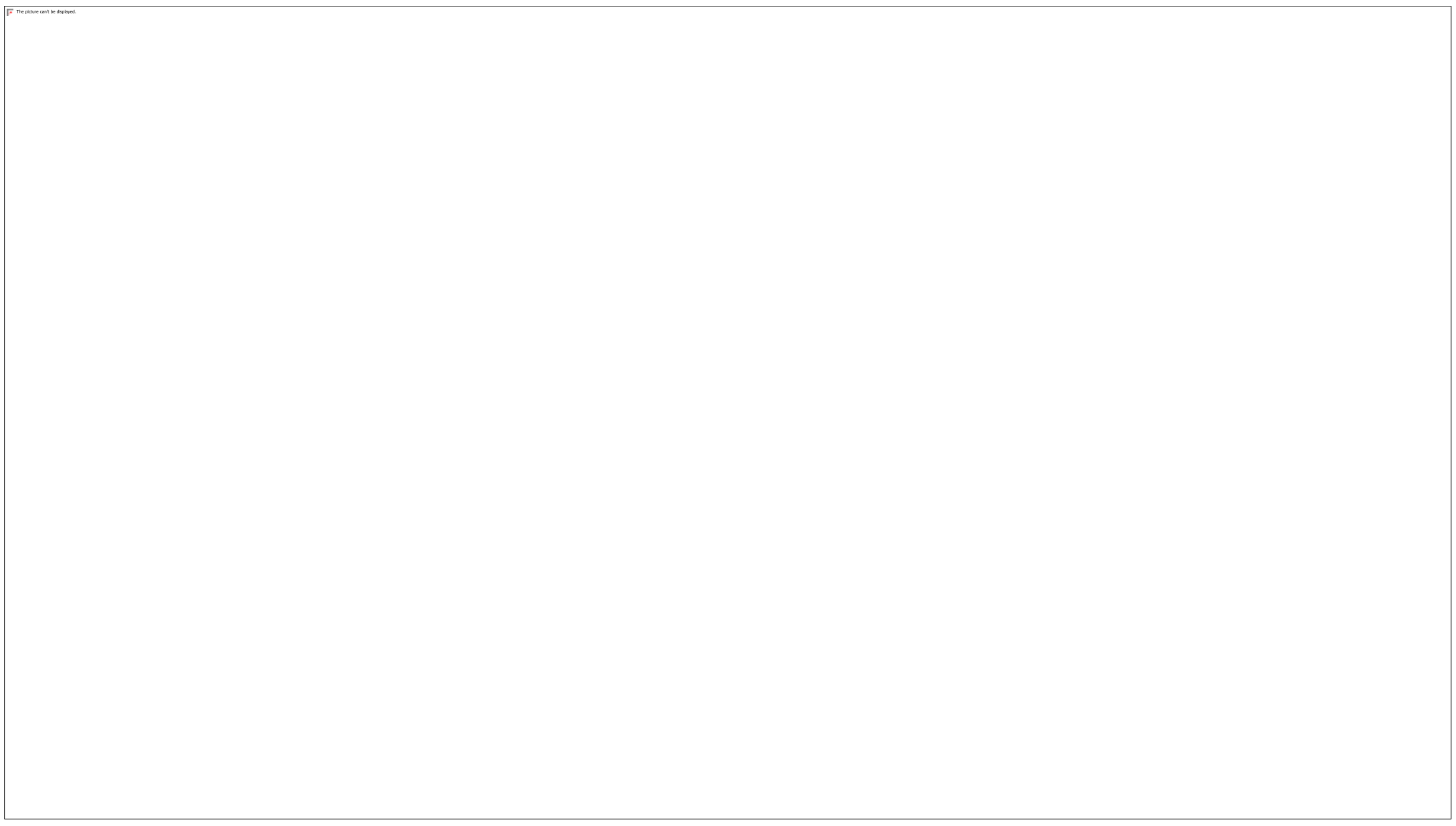

**Figure 3.** Digital Twin Data Pipeline

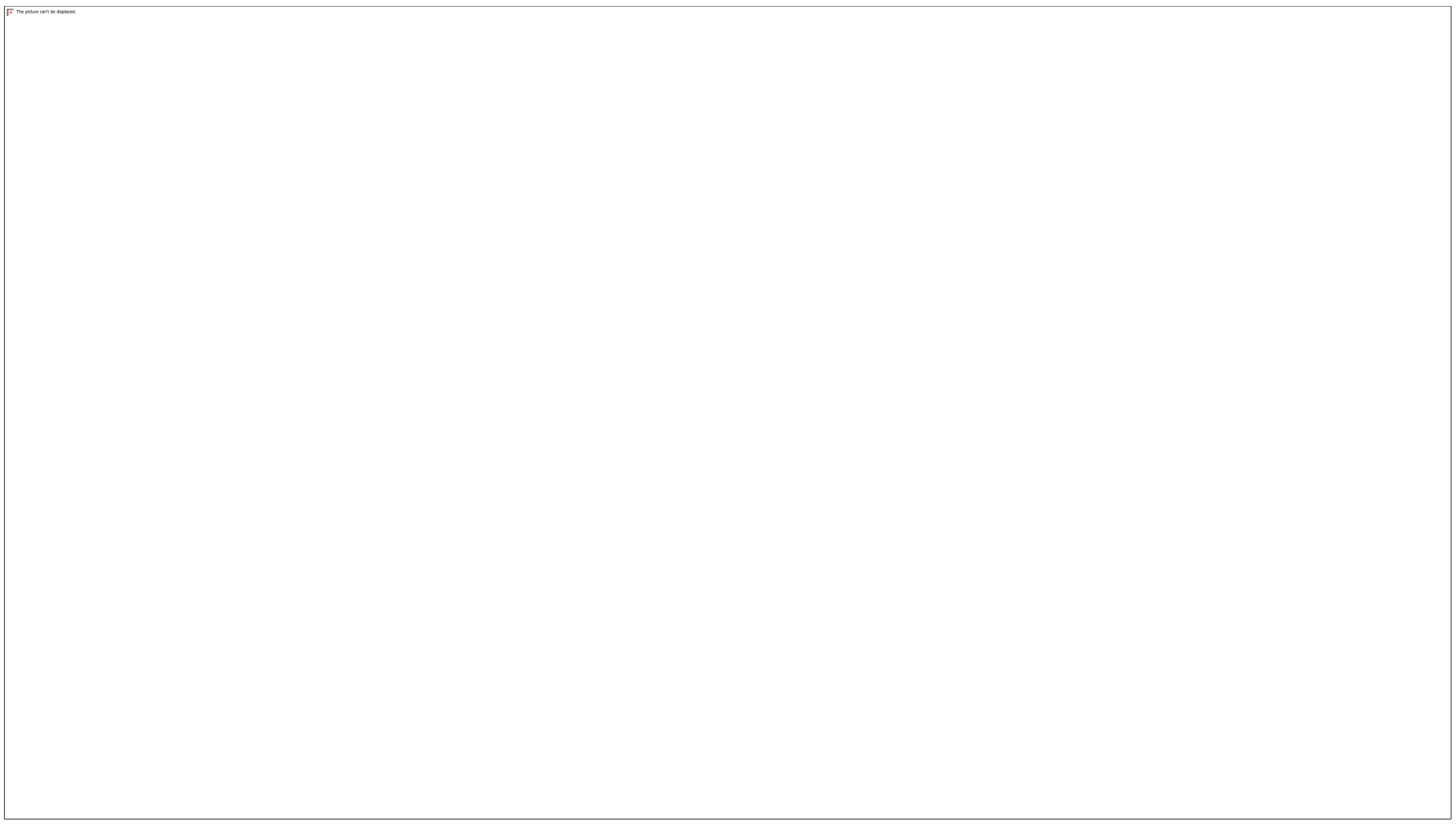

**Figure 4.** Overall architecture of the geometry pipeline. The "Reduce" steps highlighted are expanded upon in Figure 5.

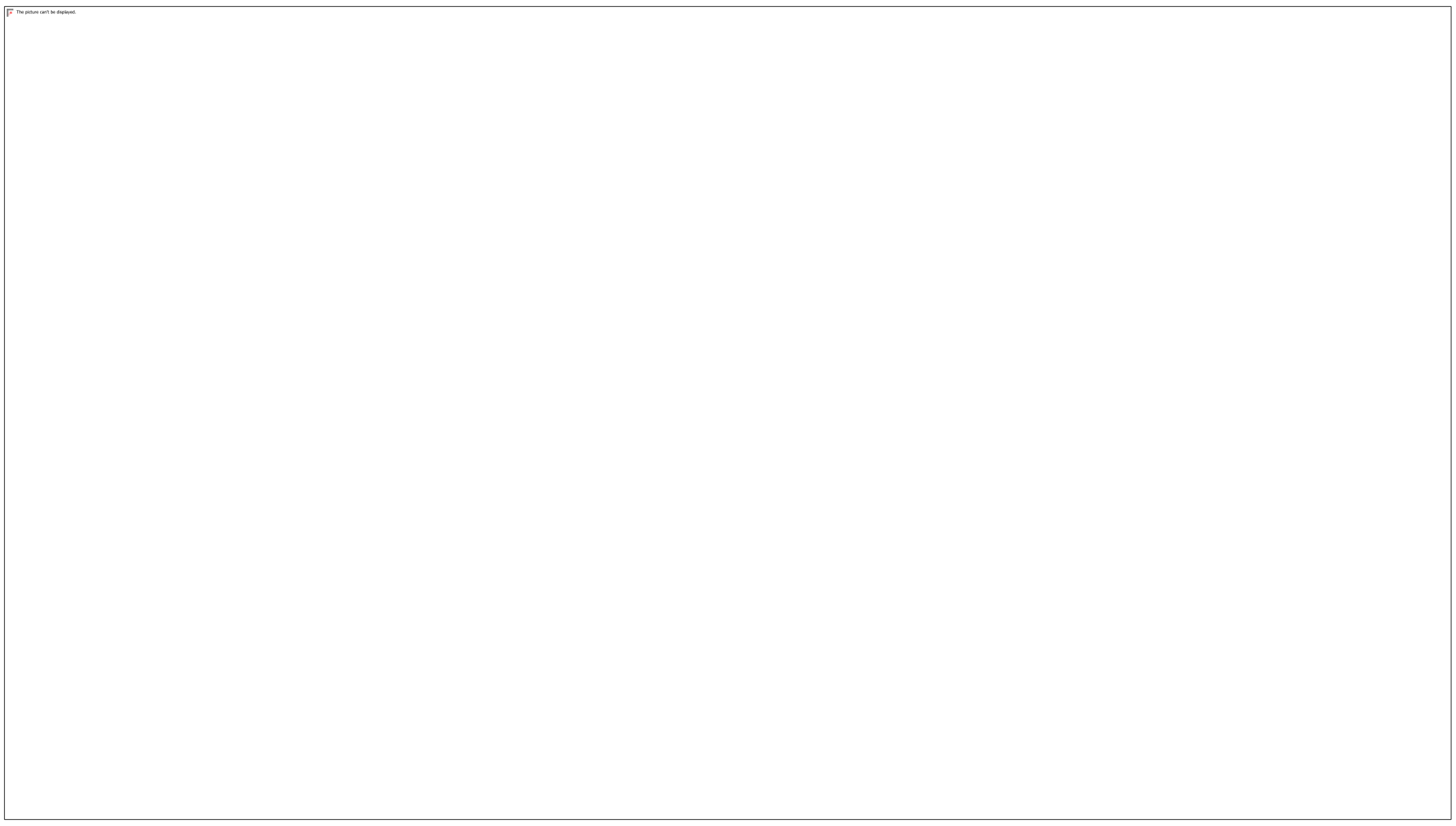

**Figure 5**. Geometry Pipeline "Reduce" step. Copies of this process are run in parallel for every unique part in the model. The three components highlighted in blue are the three machine learning models.

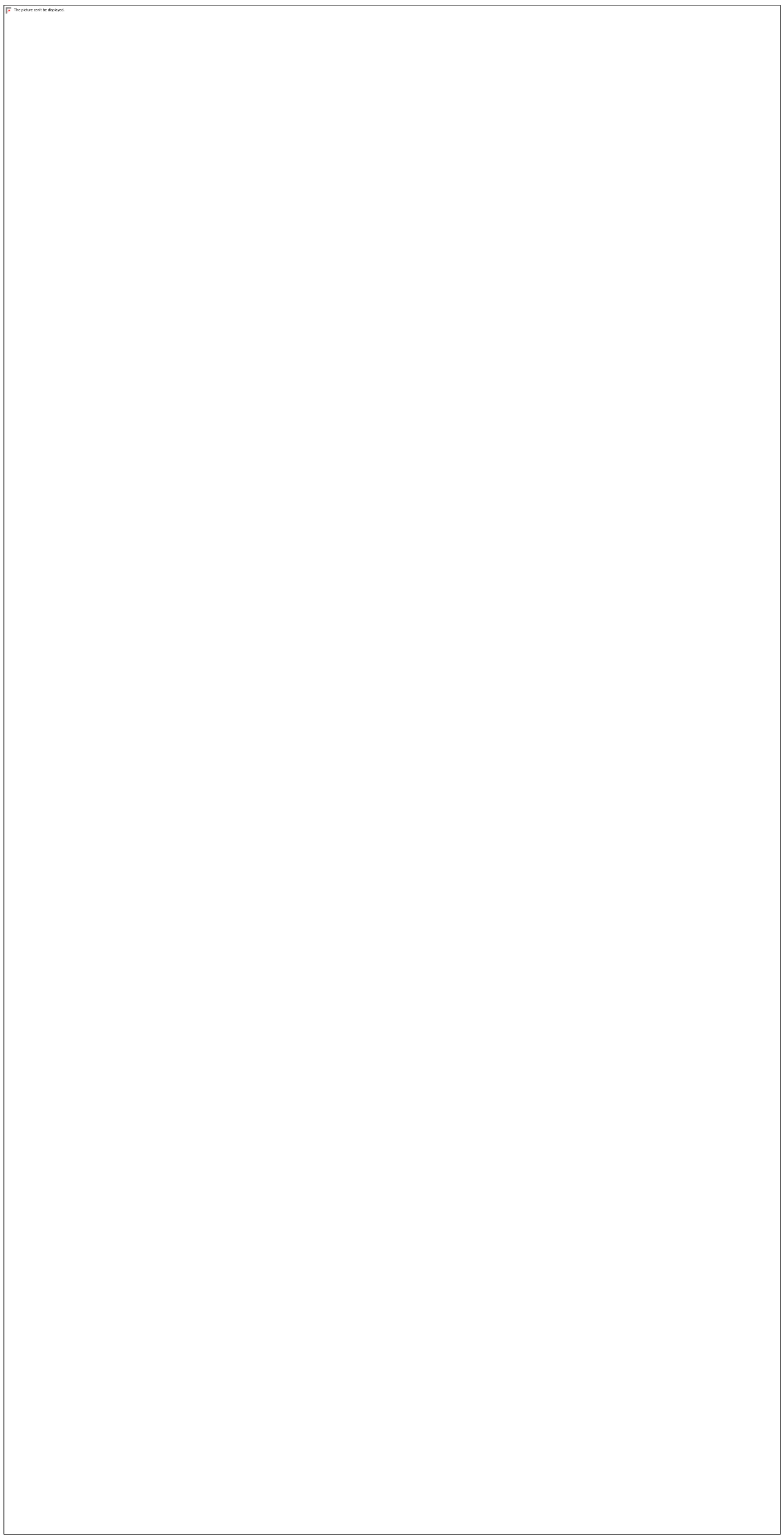

**Figure 6.** Mesh quality training data collection application (original on the left, processed on the right).

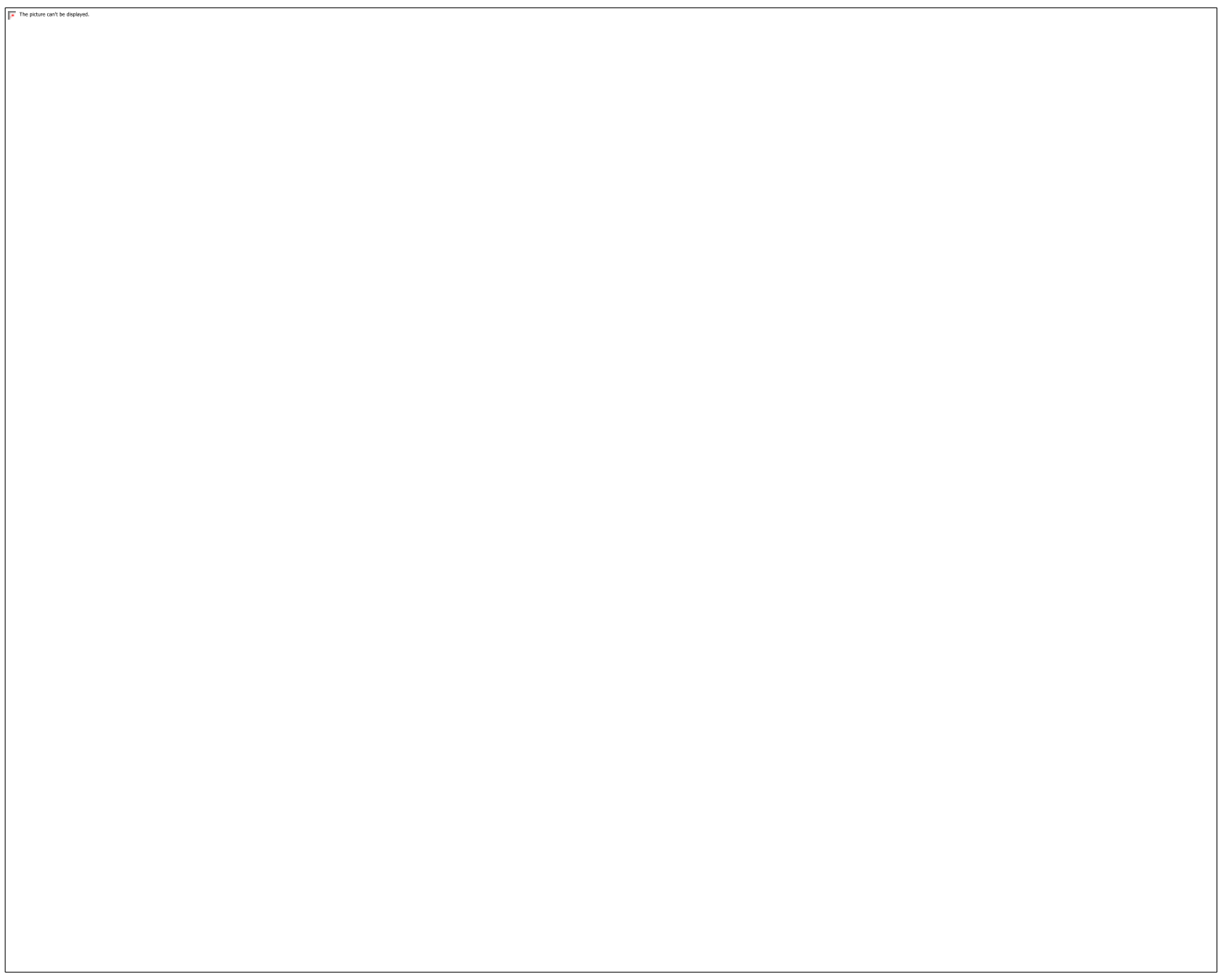

**Figure 7**. Example model produced with the geometry pipeline. The casing has been rendered transparent to show the internal details.

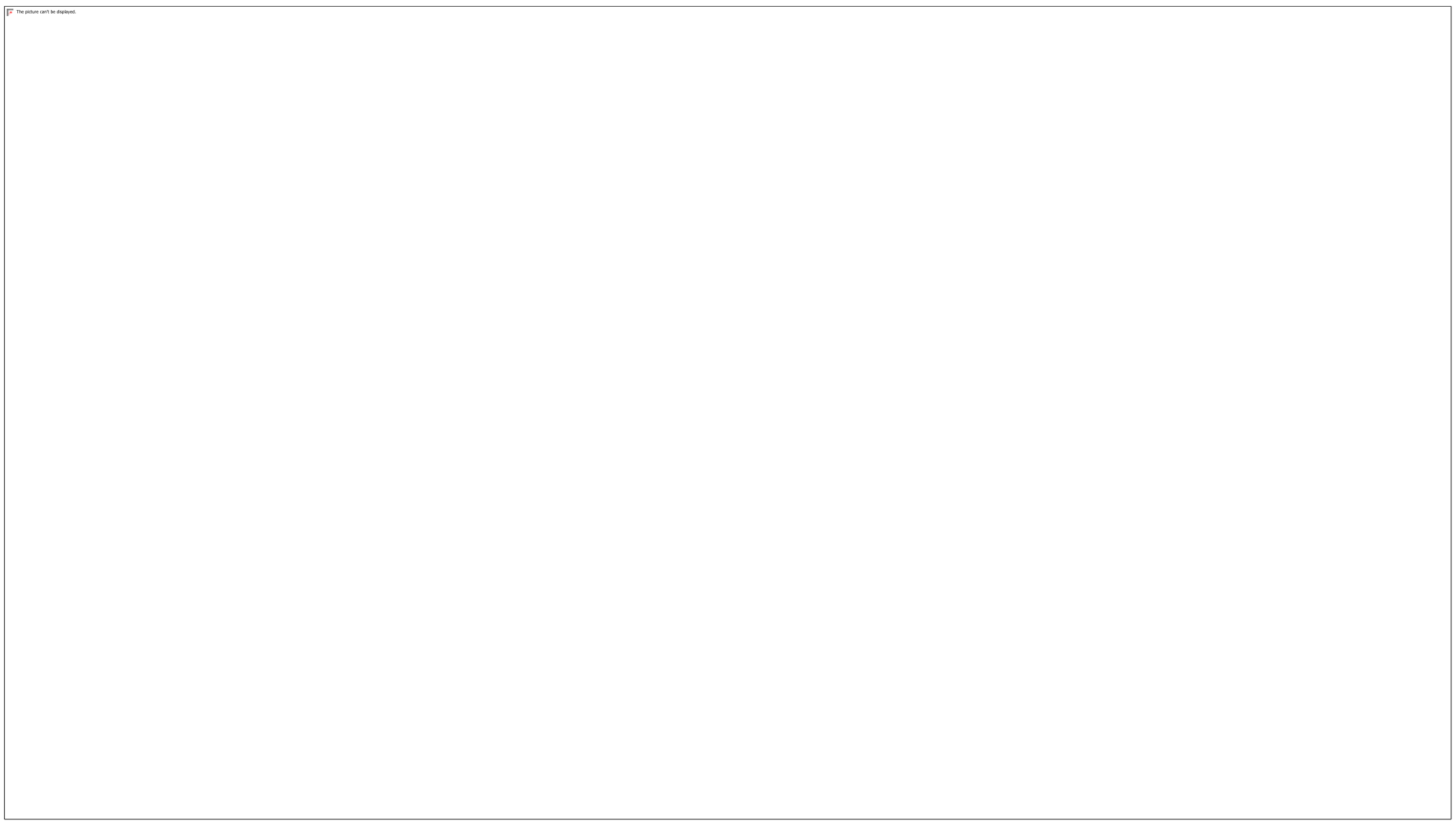

**Figure 8.** Overview of Process Pipeline Model.

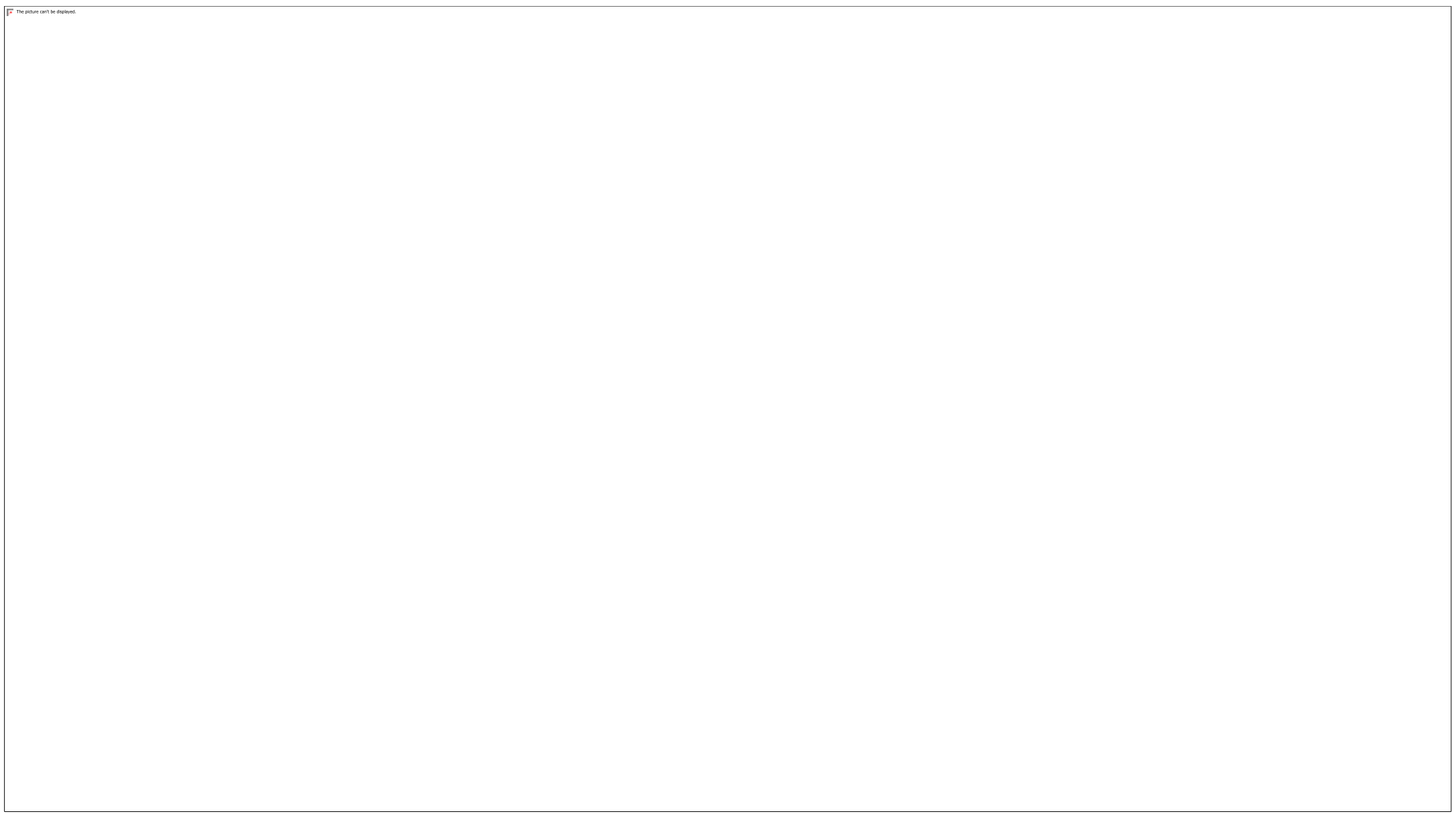

**Figure 9.** Hierarchy of an FBX file.

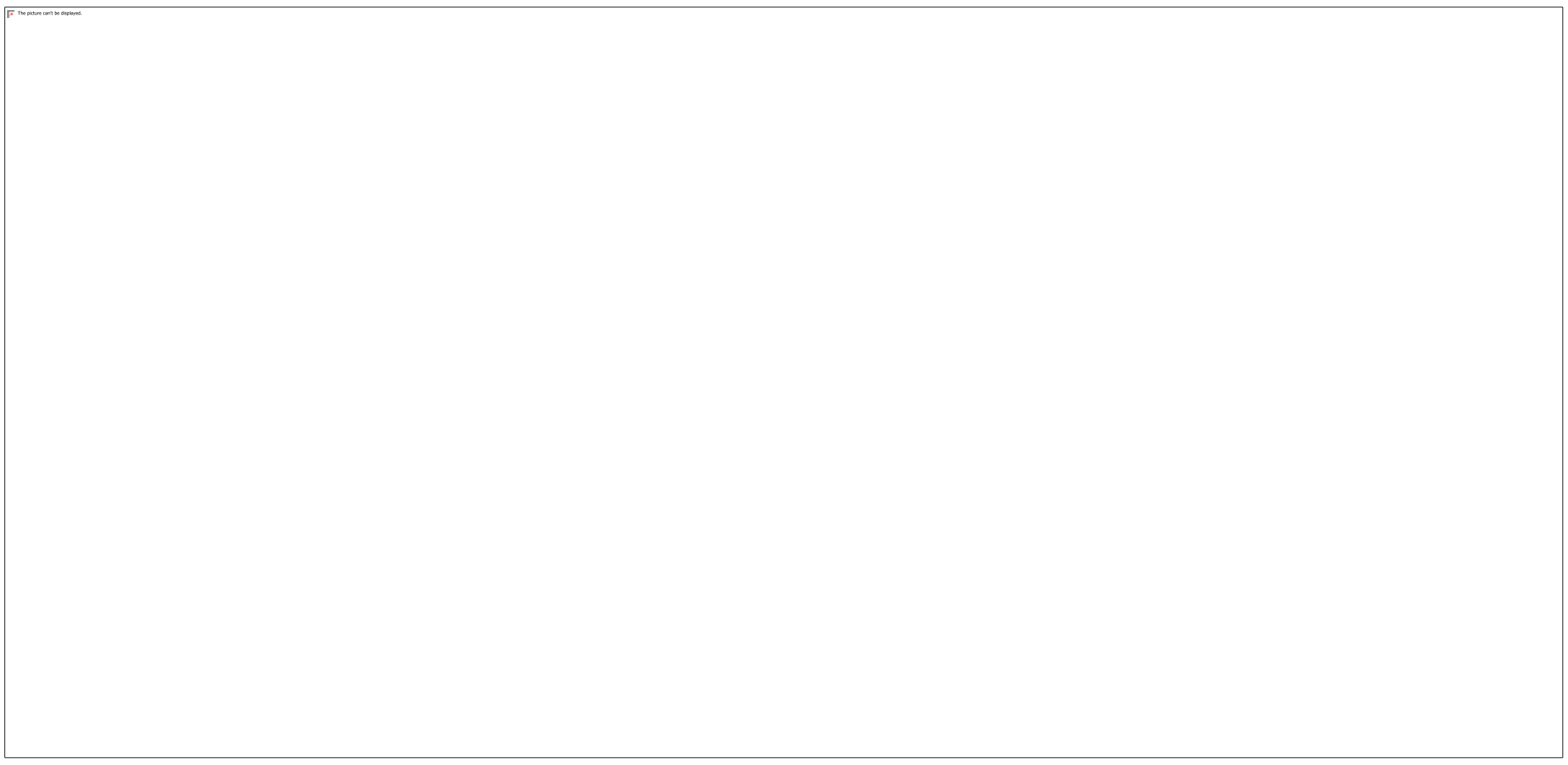

**Figure 10.** Modified FBX hierarchy used within the scenario definition.

**Figure 11**. The Scenario Definition File Link between the Digital Twin Builder and Player.

Objective: To measure the value of the armature resistance of a DC Motor.

The following procedure to be simulated to allow students to perform the laboratory tutorial.

1. Connect the circuit shown opposite.

2. Switch on the DC supply after checking all connections.

3. Increase the supply input voltage slowly from 5V to 40V with step equal to 5V.

4. Record the readings of the measurements in a table.

**Figure 12.** The Case Study Brief.

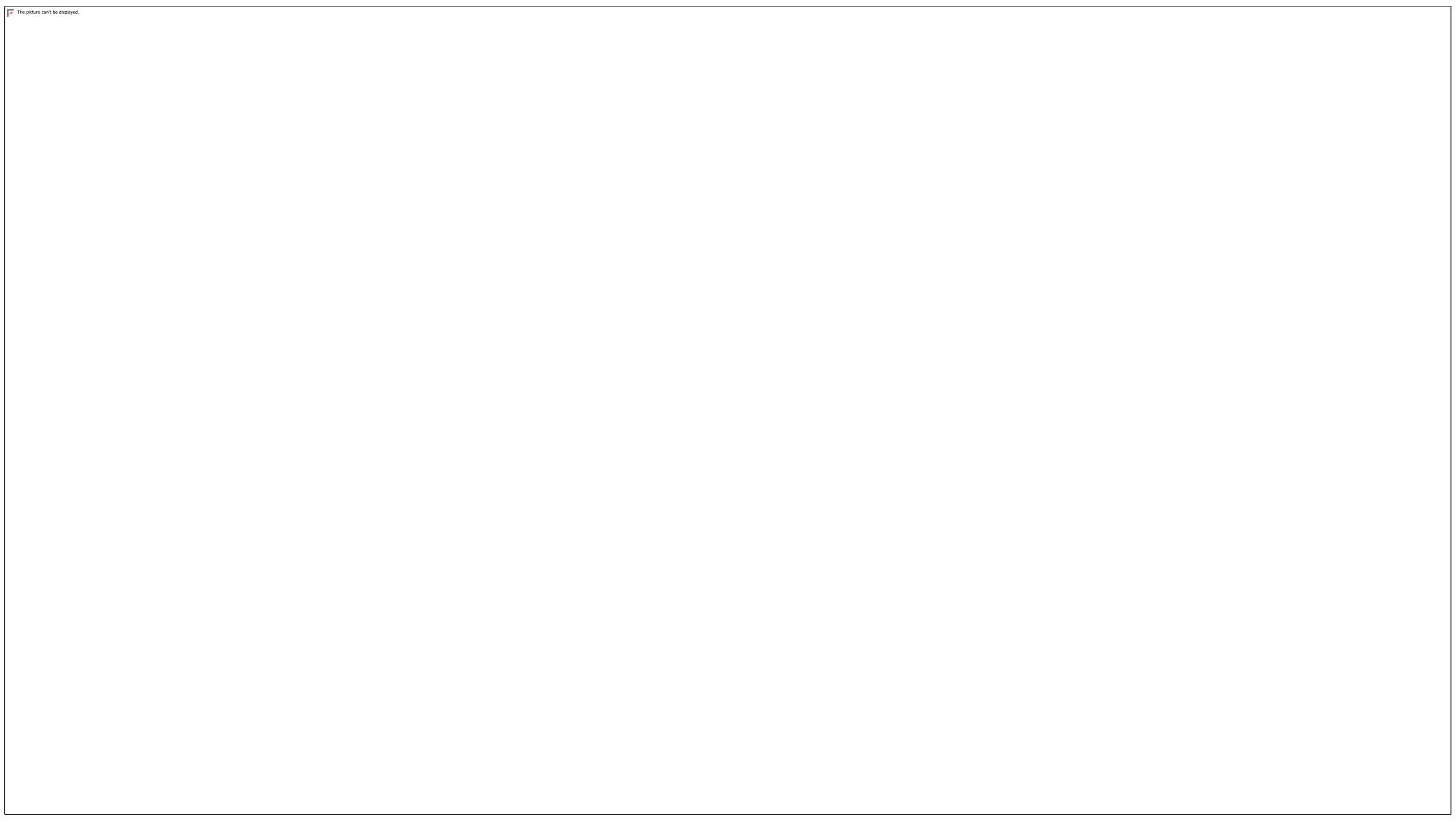

**Figure 13.** Specifying the Case Study within the Digital Twin Builder.

The picture can't be displayed.

**Figure 14.** The Electrical Laboratory within the Digital Twin Player

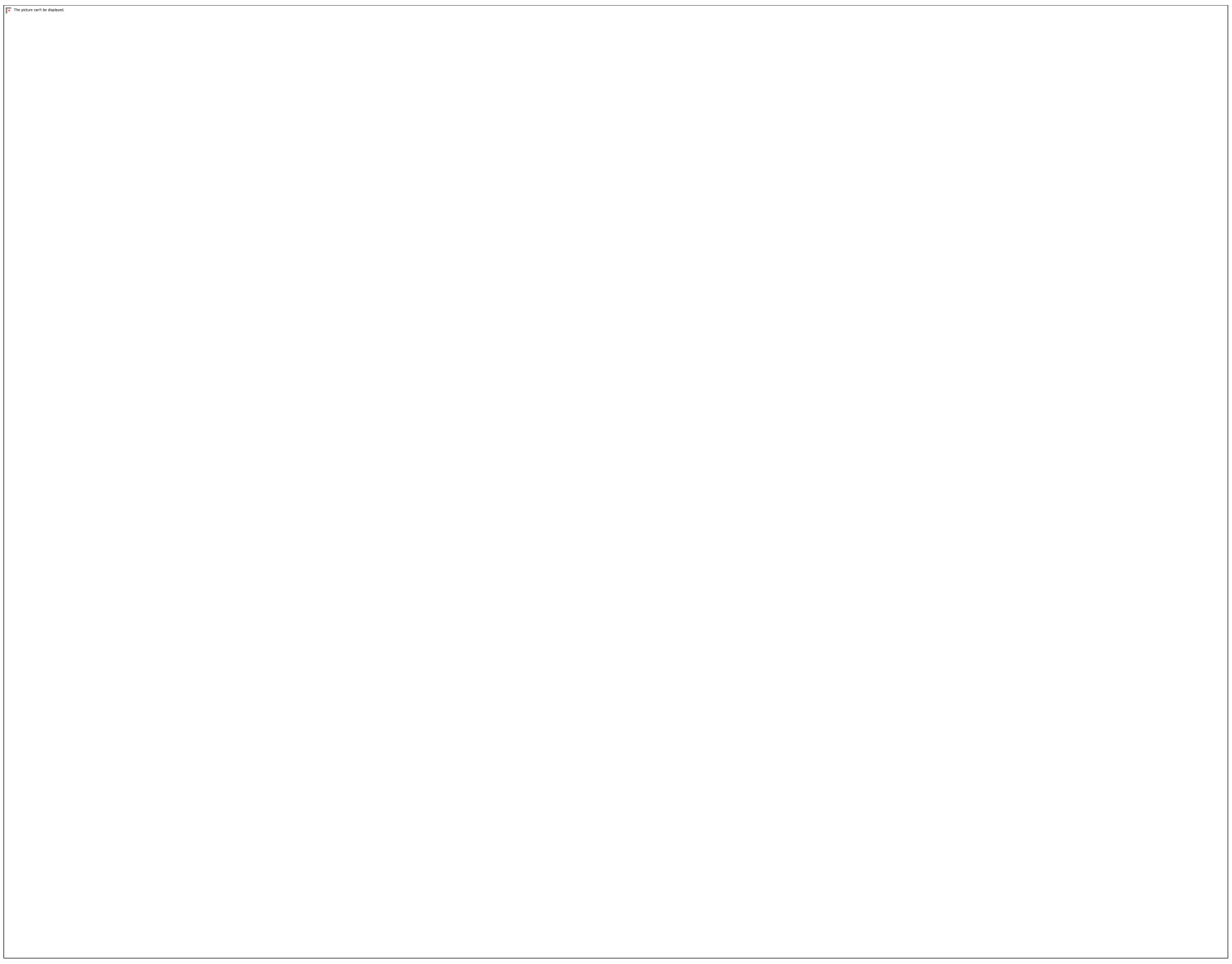

**Figure 15.** The Digital Twin Player showing an Equipment Item together with with its data simulation results.

**List of Figure Captions**

**Figure 14.** The Electrical Laboratory within the Digital Twin Player

**Figure 15.** The Digital Twin Player showing an Equipment Item together with with its data simulation results.